\documentclass[10pt,conference]{IEEEtran}
\usepackage{todonotes}
\usepackage{listings}
\usepackage{tcolorbox}
\usepackage{booktabs}
\usepackage[justification=centering]{subfig}
\usepackage{float}
\usepackage{caption}
\usepackage{balance}
\usepackage{url}
\usepackage{multirow}
\usepackage{array}
\usepackage{makecell}
\usepackage{float}
\usepackage{hyperref}
\hypersetup{hidelinks}
\usepackage[marginal]{footmisc}
\IEEEoverridecommandlockouts
\usepackage{cite}
\usepackage{amsmath,amssymb,amsfonts}
\usepackage{algorithmic}
\usepackage{graphicx}
\usepackage{textcomp}
\usepackage{xcolor}
\usepackage{booktabs}
\usepackage{enumerate}
\usepackage{enumitem}
\usepackage{threeparttable}

\newtcolorbox{qoutebox}[3][]
{
  colframe=boxcol!30!white,
  colback  = #2!10,
  #1,
}

\begin{document}

\title{Understanding Code Smell Detection via Code Review: A Study of the OpenStack Community}

\author{
    \IEEEauthorblockN{Xiaofeng Han$^{1}$, Amjed Tahir$^{2}$, Peng Liang$^{1*}$\thanks{\indent This work was partially funded by the National Key R\&D Program of China with Grant No. 2018YFB1402800.}, Steve Counsell$^{3}$, Yajing Luo$^{1}$}
    \IEEEauthorblockA{$^1$ School of Computer Science, Wuhan University, Wuhan, China}
    \IEEEauthorblockA{$^2$ School of Fundamental Sciences, Massey University, Palmerston North, New Zealand}
    \IEEEauthorblockA{$^3$ Department of Computer Science, Brunel University London, London, United Kingdom}
    \IEEEauthorblockA{hanxiaofeng@whu.edu.cn, a.tahir@massey.ac.nz, liangp@whu.edu.cn, steve.counsell@brunel.ac.uk, luoyajing@whu.edu.cn}
}

\maketitle

\begin{abstract}
Code review plays an important role in software quality control. A typical review process would involve a careful check of a piece of code in an attempt to find defects and other quality issues/violations. One type of issues that may impact the quality of the software is code smells - i.e., bad programming practices that may lead to defects or maintenance issues. Yet, little is known about the extent to which code smells are identified during code reviews. To investigate the concept behind code smells identified in code reviews and what actions reviewers suggest and developers take in response to the identified smells, we conducted an empirical study of code smells in code reviews using the two most active OpenStack projects (Nova and Neutron). We manually checked 19,146 review comments obtained by keywords search and random selection, and got 1,190 smell-related reviews to study the causes of code smells and actions taken against the identified smells. Our analysis found that 1) code smells were not commonly identified in code reviews, 2) smells were usually caused by \textit{violation of coding conventions}, 3) reviewers usually provided constructive feedback, including fixing (refactoring) recommendations to help developers remove smells, and 4) developers generally followed those recommendations and actioned the changes. Our results suggest that 1) developers should closely follow coding conventions in their projects to avoid introducing code smells, and 2) \textit{review-based} detection of code smells is perceived to be a trustworthy approach by developers, mainly because reviews are context-sensitive (as reviewers are more aware of the context of the code given that they are part of the project's development team). 
\end{abstract}


\section{Introduction}
\label{sec:introduction}

Code smells are identified as symptoms of possible code or design problems \cite{martin2018refactoring}, which may potentially have a negative impact on software quality, such as maintainability \cite{palomba2018@maintainability}, code readability \cite{Abbes2011}, testability \cite{Tahir2016}, and defect-proneness \cite{Khomh2009}.

A large number of studies have focused on smell detection and removal techniques (e.g., \cite{Tsantalis2009,moha2009decor}) with many static analysis tools for smell detection; these include tools such as PMD\footnote{\url{https://pmd.github.io}}, SonarQube\footnote{\url{https://www.sonarqube.org}}, and Designite\footnote{\url{https://www.designite-tools.com}}. 
However, previous work \cite{Yamashita2013e,tahir2020stackexchange} indicated that the program context and domain are important in identifying smells. This makes it difficult for program analysis tools to correctly identify smells since contextual information is rarely taken into account. Existing smell detection tools are also known to produce false positives \cite{Fontana2016,sharma2018survey}; therefore, manual detection of smells could be considered more valuable than automatic approaches.

Code review is a process which aims to verify the quality of the software by detecting defects and other issues in the code, and to ensure that the code is readable, understandable, and maintainable. It has been linked to improved quality \cite{baker1997code}, reduced defects \cite{mcintosh2016empirical}, reduced anti-patterns \cite{morales2015code}, and the identification of vulnerabilities \cite{meneely2014empirical}. Compared to smell detection static analysis tools, code reviews are usually performed by developers belonging to the same project \cite{mcconnell2004code}, so it is possible that reviewers will take full account of contextual information and thus better identify code smells in the code.

However, little is known about the extent to which code smells are identified during code reviews, and whether developers (the code authors) take any action when a piece of code is deemed ``smelly'' by reviewers. Therefore, we set out to study the concept behind code smells identified in code reviews and track down actions taken after reviews were carried out. To this end, we mined code review discussions from the two most active OpenStack\footnote{\url{https://www.openstack.org}} projects: Nova\footnote{\url{https://wiki.openstack.org/wiki/Nova}} and Neutron\footnote{\url{https://wiki.openstack.org/wiki/Neutron}}. We then conducted a quantitative and qualitative analysis to study how common it was for reviewers to identified code smells during code review, why the code smells were introduced, what actions they recommended for those smells, and how developers proceeded with those recommendations. In total, we analyzed 1,190 smell-related reviews got by manually checking 19,146 review comments to achieve our goal.

Our results suggest that: 1) code smells are not widely identified in modern code reviews, 2) following coding conventions can help reducing the introduction of code smells, 3) reviewers usually provide useful suggestions to help developers better fix the identified smells; while developers commonly accept reviewers' recommendations regarding the identified smells and they tend to refactor their code based on those recommendations, and 4) \textit{review-based} detection of code smells is seen as a trustworthy mechanism by developers.

The paper is structured as follows: related work is presented in Section \ref{sec:relatedWork}, the study design and data extraction methods are explained in Section \ref{sec:methodology}, the results are presented in Section \ref{sec:results}, followed by a discussion of the results in Section \ref{sec:discussion}, and the threats to the validity of the results are covered in Section \ref{sec:threats}, followed by conclusions and future work in Section \ref{sec:conclusion}.
\section{Related Work}
\label{sec:relatedWork}

\subsection{Studies on Code Smells}
A growing number of studies have investigated the impact of code smells on software quality, including defects \cite{Hall2014,Khomh2009}, maintenance \cite{Sjoberg2013}, and program comprehension \cite{Abbes2011}. Other studies have looked at the impact of code smells on software quality using a group of developers working on a specific project \cite{Sjoberg2013,Palomba2014,Soh2016}. 

Tufano \textit{et al.} \cite{tufano2015and} mined version histories of 200 open source projects to study when code smells were introduced and the main reason behind their interaction. It was found that smells appeared in general as a result of maintenance and evolution activities. Sj{\^o}berg \textit{et al.} \cite{Sjoberg2013} investigated the relationship between the presence of code smells and maintenance effort through a set of control experiments. Their study did not find significant evidence that the presence of smells led to increased maintenance effort. Previous studies also include work investigating the impact of different forms of smells on software quality, such as architectural smells \cite{garcia2009,martini2018identifying}, test smells \cite{Bavota2015,Tahir2016}, and spreadsheet smells \cite{Dou2014smells}.

A number of previous studies have investigated developer's perception of code smells and their impact in practice.
A survey on developers' perception of code smells conducted by Palomba \textit{et al.} \cite{Palomba2014b} found that developer experience and system knowledge are critical factors for the identification of code smells. Yamashita and Moonen \cite{Yamashita2013e} reported that developers are moderately concerned about code smells in their code. A more recent study by Taibi \textit{et al.} \cite{taibi2017developers} replicated the two previous studies \cite{Yamashita2013e,Palomba2014b} and found that the majority of developers always considered smells to be harmful; however, it was found that developers perceived smells as critical in theory, but not as much in practice. Tahir \textit{et al.} \cite{tahir2020stackexchange} mined posts from Stack Exchange sites to explore how the topics of code smells and anti-patterns were discussed amongst developers. Their study found that developers widely used online forums to ask for general assessments of code smells or anti-patterns instead of asking for particular refactoring solutions.

\subsection{Code Reviews in Software Development}
Code review is an integral part in modern software development. In recent years, empirical studies on code reviews have investigated the potential code review factors that affect software quality. For example, McIntosh \textit{et al.} \cite{McIntosh2014impact} investigated the impact of code review coverage and participation on software quality in the Qt, VTK, and ITK projects. The authors used the incidence rates of post-release defects as an indicator and found that poorly reviewed code (e.g. with low review coverage and participation) had a negative impact on software quality. 
A study by Kemerer \textit{et al.} \cite{kemerer2009impact} investigated the impact of review rate on software quality. The authors found that the \textit{Personal Software Process} review rate was a significant factor affecting defect removal effectiveness, even after accounting for developer ability and other significant process variables.

Several studies \cite{mcintosh2016empirical, McIntosh2014impact, Kononenko2015investigating} have investigated the impact of modern code review on software quality. Other studies have also investigated the impact of code reviews on different aspects of software quality, such as vulnerabilities \cite{bosu2014identifying}, design decisions \cite{zanaty2018empirical}, anti-patterns \cite{morales2015code}, and code smells \cite{nanthaamornphong2016empirical, pascarella2020reviews}. Aziz and Apatta \cite{nanthaamornphong2016empirical} examined review comments from code reviewers and described the need for an empirical analysis of the relationship between code smells and peer code review. Their preliminary analysis of review comments from OpenStack and WikiMedia projects indicated that code review processes identified a number of code smells. However, the study only provided preliminary results and did not investigate the causes or resolution strategies of these smells. A more recent study by Pascarella \textit{et al.} \cite{pascarella2020reviews} found that code reviews helped in reducing the severity of code smells in source code, but this was mainly a side effect to other changes unrelated to the smells themselves. 
\section{Methodology}
\label{sec:methodology}

\begin{figure*}[h]
    \centering
    \includegraphics[width=0.9\linewidth]{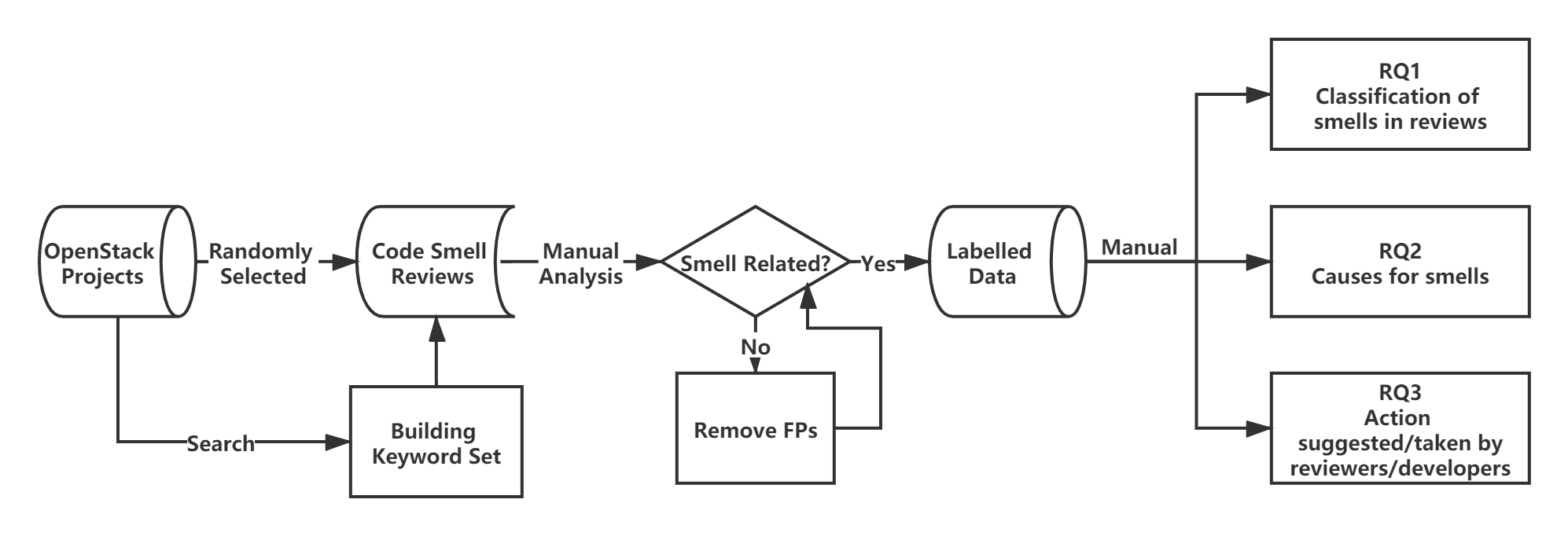}
    \caption{An overview of our data mining and analyzing process}
    \label{fig:data_mining_and_extraction}
\end{figure*}

\subsection{Research Questions}
The goal of this study is to investigate the concept behind code smells identified during code reviews and what actions are suggested by those reviewers and performed by developers in response to the identified smells. To achieve this goal, we formulated the following three research questions (RQs).\\

\noindent\textbf{RQ1: Which code smells are the most frequently identified by code reviewers?}

\noindent\textbf{Rationale}: This question aims to find out how frequent smells are identified by code reviewers and what particular code smells are repeatedly detected by reviewers. Such information can help in improving developers' awareness of these frequently identified code smells.\\ 

\noindent\textbf{RQ2: What are the common causes for code smells that are identified during code reviews?}

\noindent\textbf{Rationale}: This question investigates the main reasons behind the identified smells as explained by the reviewers or developers. When reviewing code, reviewers can express why they think the code under review may contain a smell. Developers can also reply to reviewers and explain how they introduced the smells. Understanding the common causes of smells identified manually by reviewers will shed some light on the effectiveness of manual detection of smells and help developers better understand the nature of identified smells and reduce such smells in the future.\\

\noindent\textbf{RQ3: How do reviewers and developers treat the identified code smells?}

\noindent\textbf{Rationale}: This question investigates the actions suggested by reviewers and those taken by developers on the identified smells. When a smell is identified, reviewers can provide suggestions to resolve the smell, and developers can then decide on whether to fix or ignore it. This question can be further decomposed into three sub-research questions from the perspective of reviewers, developers, and the relationship between their actions: 

\textbf{RQ3.1: What actions do reviewers \textit{suggest} to deal with the identified smells?}

\textbf{RQ3.2: What actions do developers \textit{take} to resolve the identified smells?}

\textbf{RQ3.3: What is the \textit{relationship} between the actions suggested by reviewers and those taken by developers?}

\subsection{OpenStack Projects and Gerrit Review Workflow}
\label{sec:subject_projects}

OpenStack is a set of software tools for building and managing cloud computing platforms. It is one of the largest open source communities. Based on most recent data, OpenStack projects contain around 13 million lines of code, contributed to by around 12k developers\footnote{As of October 2020: \url{https://www.openhub.net/p/openstack}}. 
We deemed the platform to be appropriate for our analysis, since the community has long invested in its code review process\footnote{\url{https://docs.opendev.org/opendev/infra-manual/latest/developers.html}}. 

We then selected two of the most active OpenStack projects as our subject projects: Nova (a fabric controller) and Neutron (a network connectivity platform) - Table \ref{tab:subject_projects} provides an overview of the data obtained from the two projects. 
Both projects are written in Python, and use Gerrit\footnote{\url{https://www.gerritcodereview.com}}, a web-based code review platform built on top of Git. The Gerrit review workflow is explained next.

\begin{table}[h]
\caption{An overview of the subject projects (Nova and Neutron)}
\label{tab:subject_projects}
\resizebox{\columnwidth}{!}{%
\begin{tabular}{@{}lccc@{}}
\toprule
\textbf{Project} & \textbf{Review Period} & \textbf{\#Code Changes} & \textbf{\#Comments} \\ \midrule
\textbf{Nova}             & Jan 14 - Dec 18        & 22,762                   & 156,882             \\
\textbf{Neutron}          & Jan 14 - Dec 18        & 15,256                   & 152,429             \\ \hline
\textbf{Total} &                 & 38,018                   & 309,311             \\ \bottomrule
\end{tabular}}
\end{table}

Gerrit provides a detailed code review workflow. First, a developer (author) makes a change to the code and submits the code (patch) to the Gerrit server so that it can be reviewed. Then, verification bots check the code using static analysers and run automated tests. A reviewer (usually other developers that have not been involved in writing the code under review) will then conduct a formal review of the code and provide comments. The original author can reply to the reviewer's comments and action the required changes by producing a new revision of the patch. This process is repeated until the change is merged to the code base or abandoned by the author.

\subsection{Mining Code Review Repositories}
\label{mining_code_reivew_repositories}
Fig. \ref{fig:data_mining_and_extraction} outlines our data extraction and mining process. 
We mined code review data {\em via} the RESTful API provided by Gerrit, which returns the results in a JSON format. We used a Python script to automatically mine the review data in the studied period and store the data in a local database. In total, we mined 38,018 code changes and 309,311 review comments between Jan 2014 and Dec 2018 from the two projects.

\subsection{Building the Keyword Set}
\label{build_keyword_set}
To locate code review comments that include code smell discussions, we used several variations of terms referring to code smells or anti-patterns, including ``code smell'', ``bad smell'', ``bad pattern'', ``anti-pattern'', and ``technical debt''. In addition, considering that reviewers may point out the specific code smell by its name (e.g., dead code) rather than using generic terms, we included a list of code smell terms obtained from Tahir \textit{et al.} \cite{Tahir2018EASE}, that extracted these smell terms from several relevant studies on this topic, including the first work on code smells by Fowler \cite{martin2018refactoring} and the systematic review by Zhang \textit{et al.} \cite{zhang2011code}.
The list of smell terms used in our study are shown in Table \ref{tab:terms}.

\begin{table*}[h]
\centering
\caption{Code smell terms included in our mining}
\resizebox{0.8\linewidth}{!}{
\begin{tabular}{@{}llll@{}}
\toprule
\multicolumn{4}{c}{\textbf{Code Smell Terms}}                                                                           \\ \toprule
Accidental Complexity               & Anti Singleton                    & Bad Naming                & Blob Class                \\
Circular Dependency                 & Coding by Exception               & Complex Class             & Complex Conditionals      \\
Data Class                          & Data Clumps                       & Dead Code                 & Divergent Change          \\
Duplicated Code                     & Error Hiding                      & Feature Envy              & Functional Decomposition  \\
God Class                           & God Method                        & Inappropriate Intimacy    & Incomplete Library Class  \\
ISP Violation                       & Large Class                       & Lazy Class                & Long Method               \\
Long Parameter List                 & Message Chain                     & Middle Man                & Misplaced Class           \\
Parallel Inheritance Hierarchies    & Primitive Obsession               & Refused Bequest           & Shotgun Surgery           \\
Similar Subclasses                  & Softcode                          & Spaghetti Code            & Speculative Generality    \\
Suboptimal Information Hiding       & Swiss Army Knife                  & Temporary Field           & Use Deprecated Components \\ \bottomrule
\end{tabular}
}
\label{tab:terms}
\end{table*}

Since the effectiveness of keyword-based mining approach relies on the set of keywords that are used in the search, we followed the systematic approach used by Bosu \textit{et al.} \cite{bosu2014identifying} to identify the keywords included in our search. This includes the following steps\footnote{implemented using the NLTK package: \url{http://www.nltk.org}}:

\begin{enumerate}
    \item Build an initial keyword set.
    \item Build a corpus by searching for review comments that contain at least one keyword of our initial keyword set (e.g., ``dead'' or ``duplicated'') in the code review data we collected in Section \ref{mining_code_reivew_repositories}.
    \item Process the identified review comments which contain at least one keyword of our initial keyword set, and then apply the identifier splitting rules (i.e., ``isDone'' becomes ``is Done'' or ``is\_done" becomes ``is done").
    \item Create a list of tokens for each document in the corpus.
    \item Clean the corpus by removing stopwords, punctuation, and numbers, and then convert all the words to lowercase.
    \item Apply the Porter stemming algorithm \cite{Porter2001Snowball} to obtain the stem of each token.
    \item Create a Document-Term matrix \cite{tan2016introduction} from the corpus.
    \item Find the additional words co-occurred frequently with each of our initial keywords (co-occurrence probability of 0.05 in the same document).
\end{enumerate}

After performing these eight steps, we found that no additional keywords co-occurred with each of our initial keywords, based on the co-occurrence probability of 0.05 in the same document. Therefore, we believe that our initial keyword set is sufficient to support the keyword-based mining method.
Due to space constraints, we provide the initial set of keywords (which are the same as the final set of keywords) associated with code smells, in our replication package \cite{anonymous_replication_package}.

\subsection{Identifying Smell-related Reviews in Keywords-searched Review Comments}
\label{keywords_searched_review_comments}

We followed the following four steps to identify smell-related reviews:

In \textbf{step one}, we developed a Python script to search for review comments that contained at least one of the keywords identified in Section \ref{build_keyword_set}. The search returned a total of 18,082 review comments from the two projects.

In \textbf{step two}, to increase our verification process, two of the authors independently and manually analyzed the review comments obtained in \textbf{step one} to exclude comments  \emph{clearly} unrelated to code smells.
If a review comment was deemed by both coders to be unrelated to code smells, it was then excluded. As a result of this step, the number of review comments was reduced to 3,666.

To illustrate this process, consider the following two review comments that contain the keyword ``dead''. In the first example, the reviewer commented that ``\textit{why not to put the port on dead vlan first?}''\footnote{\url{https://review.opendev.org/c/openstack/neutron/+/179314}}. Although this comment contains the keyword ``dead'', both coders thought that it was unrelated to code smells, and the comment was therefore excluded. In the second example, the reviewer commented ``\textit{remove dead code}''\footnote{\url{https://review.opendev.org/c/openstack/neutron/+/196893}}, which was regarded  as related to ``dead code'' by the two coders and thus was included in the analysis.

In \textbf{step three}, two of the authors worked together to further manually analyze the remaining review comments.
The same two authors carefully analyzed the contextual information of each review comment, including the code review discussions and associated source code to determine whether the code reviewers identified any smells in the review comments. We considered a comment to be related to code smell only when both coders agreed. The agreement between the two authors was calculated using Cohen's Kappa coefficient \cite{jacob1960coefficient}, which was 0.85. When the coders were unsure or disagreed, a third author was then involved in the discussion until an agreement was reached. This resulted in a reduction in the number of review comments to 1,235.

\begin{figure}[htb]
    \centering
    \captionsetup{justification=centering}
    \includegraphics[width=0.95\linewidth]{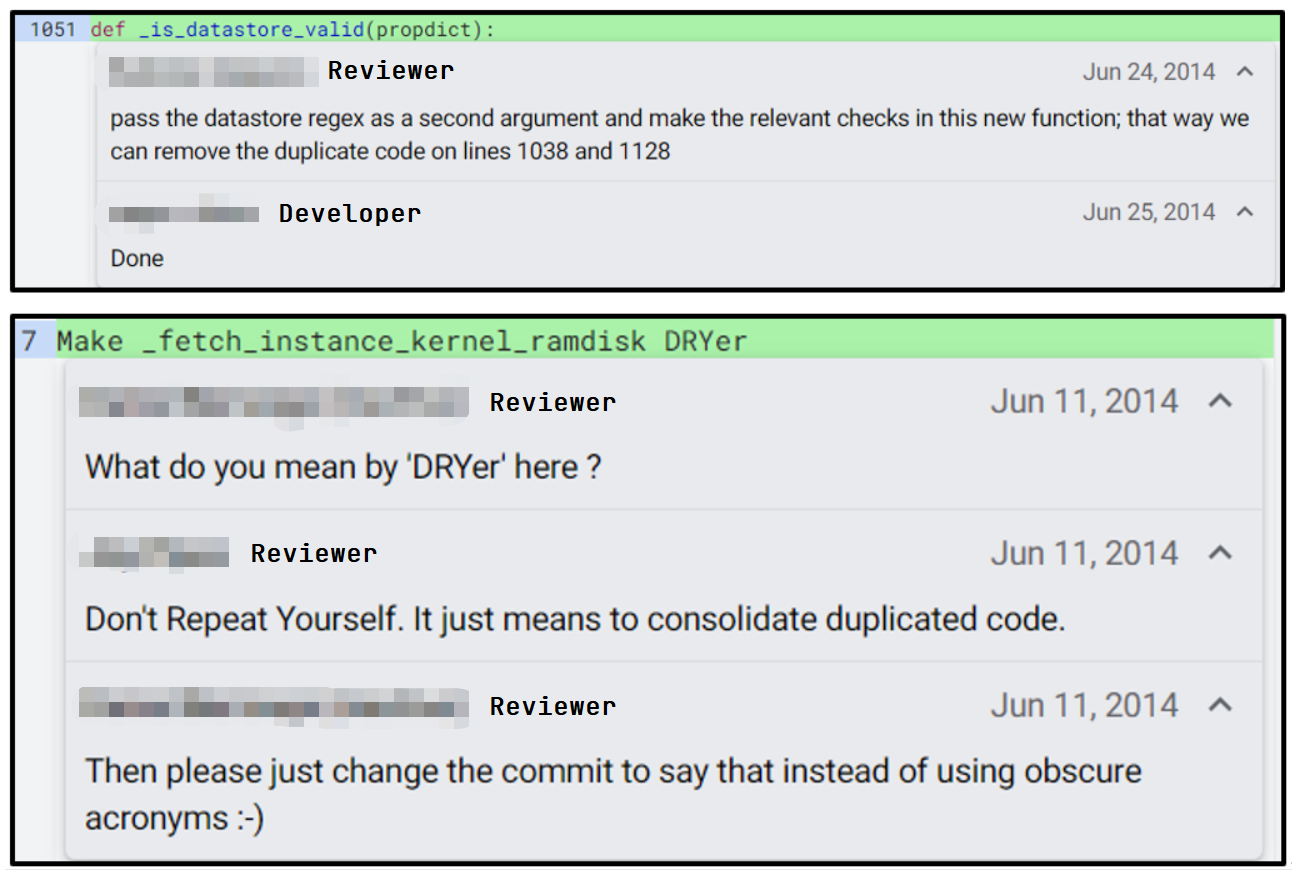}
    \caption{Review comments related to `duplicated code': the top review is smell-related, while the bottom one is not.}
    \label{fig:smell_examples}
\end{figure}

To better explain our selection process, consider the two examples in Fig. \ref{fig:smell_examples}. In the top example\footnote{\url{https://review.opendev.org/c/openstack/nova/+/100097}}, the reviewer suggested adding another argument to the method to eliminate code duplication. Then the developer replied ``\textit{Done}'', which implies an acknowledgment of the code duplication. We considered this as a clear smell-related review, and the review comment was retained for further analysis. In contrast, in the bottom example\footnote{\url{https://review.opendev.org/c/openstack/nova/+/91092}}, we observed that the comment was just used to explain the meaning of the ``DRY'' principle, but did not indicate that the code contained duplication according to the context. Thus, this comment was excluded from analysis.

Finally, in \textbf{step four}, we recorded the contextual information of each review comment in an external text file for further analysis, which contained: 1) a URL to the code change, 2) the type of the identified code smell, 3) the discussion between reviewers and developers, and 4) a URL to the source code. We ended up with a total of 1,174 smell-related reviews (we note that several review comments appearing in the same discussion were merged). An example of an extracted source file is shown below:
\\

\begin{tcolorbox}[colback = white]
\textbf{Code Change URL:} \url{http://alturl.com/2ne85}\\
\textbf{Code Smell:} \emph{Dead Code} \\
\textbf{Code Smell Discussions:} \\
\textbf{1) Reviewer:} ``Looks like copy-paste of above and, more importantly, dead code.'' \\
\textbf{2) Developer:} ``yes, sorry for that.''\\
\textbf{Source Code URL:} \url{http://alturl.com/yai68}
\end{tcolorbox}

\subsection{Identifying Smell-related Reviews in Randomly-selected Review Comments}
\label{randomly_selected_review_comments}

Knowing that reviewers and developers may not use the same keywords as we used in Section \ref{keywords_searched_review_comments} when detecting and discussing code smells during code review, we supplemented our keyword-based mining approach by including a randomly selected set of review comments from the rest of the review comments (291,229) that did not contain any of the keywords used in Section \ref{build_keyword_set}. Based on 95\% confidence level and 3\% margin of error \cite{israel1992dss}, we ended up with an additional 1,064 review comments. We then followed the same process of manual analysis (i.e., from \textbf{step two} to \textbf{step four} as described in Section \ref{keywords_searched_review_comments}) to identify smell-related reviews in these randomly selected review comments. Finally, we identified a total of 16 smell-related reviews. 

In addition to the reviews obtained by keywords search in Section \ref{keywords_searched_review_comments}, we finally obtained a total of 1,190 smell-related reviews for further analysis. We provided a full replication package containing all the data, scripts, and results online \cite{anonymous_replication_package}.

\subsection{Manual Analysis and Classification}
\label{sec:manualAnalysis}
For RQ1, in Sections \ref{keywords_searched_review_comments} and \ref{randomly_selected_review_comments}, we identified and recorded the smell type of each review when analyzing the review comments. When a reviewer used general terms (e.g., ``smelly'' or ``anti-pattern'') to describe the identified smell, we classified the type in these reviews as ``general''. The others were classified as specific smell (e.g., ``duplicated code'').

For RQ2, we adopted Thematic Analysis \cite{braun2006using} to find the causes for the identified code smells in Sections \ref{keywords_searched_review_comments} and \ref{randomly_selected_review_comments}. We used MAXQDA\footnote{\url{https://www.maxqda.com/}} - a software package for qualitative research - to code the contextual information of the identified code smells. Firstly, we coded the collected smell-related reviews by highlighting sections of the text related to the causes of the code smell in the review. When no cause was found, we used ``cause not provided/unknown''. Next, we looked over all the code that we created to identify common patterns among them and generated themes. We then reviewed the generated themes by returning to the dataset and comparing our themes against it. Finally, we named and defined each theme.
This process was performed by the same two coders in Sections \ref{keywords_searched_review_comments} and \ref{randomly_selected_review_comments}. A third author was involved in cases of disagreement by the two coders.

For RQ3, we decided to manually check the code reviews obtained in Sections \ref{keywords_searched_review_comments} and \ref{randomly_selected_review_comments} to identify the actions suggested by reviewers and taken by developers. 

For RQ3.1, we categorized the actions recommended by reviewers into three categories, which are proposed in \cite{Tahir2018EASE}:

\begin{enumerate}
    \item\textbf{Fix}: recommendations are made to refactor the code smell.
    \item \textbf{Capture}: detect that there may be a code smell, but no direct refactoring recommendations are given.
    \item \textbf{Ignore}: recommendations are to ignore the identified smells. 
\end{enumerate}

For RQ3.2, we investigated how developers responded to reviewers that identified code smells in their code. We conducted this analysis in three steps: 
We first checked the developer’s response to the reviewer in the discussion (Gerrit provides a discussion platform for both reviewers and developers). Second, we investigated the associated source code file(s) of the patch before the review was conducted, and the changes in the source code made after the review.
Finally, if the developers neither responded to reviewers nor modified source code, we then checked the status (i.e., merged or abandoned) of the corresponding code change.

We considered the identified code smells to be solved in these three cases: 1) the original developer \textit{self-admitted} a refactoring (as part of the review discussion), 2) changes were made in the source code file(s), and 3) the corresponding code change was then abandoned.

For RQ3.3, based on the results of RQ3.1 and RQ3.2, we categorized the relationship between the actions recommended by reviewers and those taken by developers into the following three categories: 

\begin{enumerate}
    \item A developer \textit{agreed} with the reviewer’s recommendations.
    \item A developer \textit{disagreed} with the reviewer’s recommendations, or
    \item A developer \textit{did not respond} to the reviewer’s comments.
\end{enumerate}

These three categories were then mapped into two actions: 1) \textit{fixed the smell} (i.e., refactoring was done) or 2) \textit{ignored the change} (i.e., no changes were performed to the source code with regard to the smell).

This process was conducted by the first author and the result of each step was cross-validated by another author. Again, a third author was involved in case of disagreement.
In total, the manual analysis process took around thirty days of full-time work of the coders. We also provided a full replication package containing all the data, scripts, and results from the manual analysis online \cite{anonymous_replication_package}.
\section{Results}
\label{sec:results}

In this section, we present the results of our three RQs. We note that, due to space constraints, detailed results of our analysis are provided externally \cite{anonymous_replication_package}.

\textbf{RQ1: Which code smells are the most frequently identified by code reviewers?}

\begin{figure}[h]
    \centering
    \includegraphics[width=\linewidth]{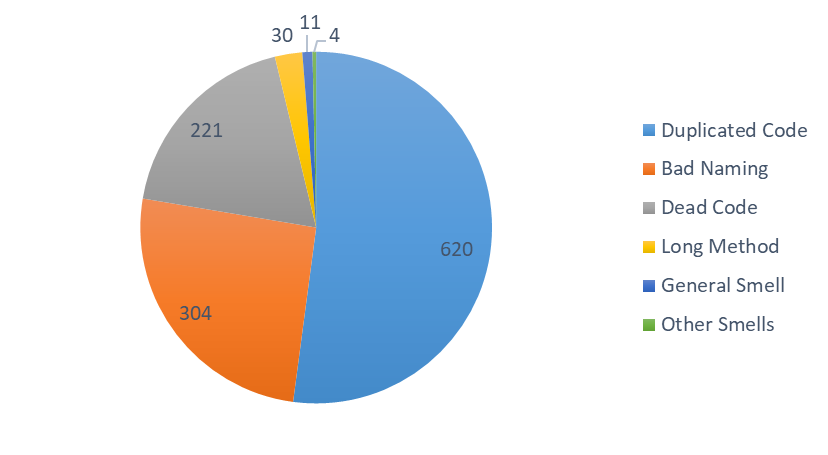}
    \caption{Number of reviews for the identified code smells}
    \label{fig:discussions_for_specific_term}
\end{figure}

Figure \ref{fig:discussions_for_specific_term} shows the distribution of code smells identified in the code reviews obtained in Sections \ref{keywords_searched_review_comments} and \ref{randomly_selected_review_comments}. In general, we identified 1,190 smell-related reviews. Compared to the number of all the review comments we obtained, we found that code smells are not commonly identified in code reviews. In addition, of all the code smells we identified, \emph{duplicated code} is by far the most frequently identified smell by name, with exactly 620 instances. The smells of \emph{bad naming} and \emph{dead code} were also frequently identified, as they were discussed in 304 and 221 code reviews, respectively. There were 30 code reviews which identified \emph{long method}, while other smells such as \emph{circular dependency} and \emph{swiss army knife} were discussed in only 4 code reviews. The rest of code reviews (11) used general terms (e.g., code smell) to describe the identified smells.

\noindent\resizebox{\columnwidth}{!}{\fbox{
	\parbox{\columnwidth}{
		\textbf{RQ1: the most frequently identified smells in code reviews.}
		\\
	Code smells are not widely identified in code reviews. Of the identified smells, \textbf{\emph{duplicated code}}, \textbf{\emph{bad naming}}, and \textbf{\emph{dead code}} are the most frequently identified smells in code reviews.
	}
}}
\\

\textbf{RQ2: What are the common causes for code smells that are identified during code reviews?}

For RQ2, we used Thematic Analysis to identify the common causes for the identified code smells as noted by code reviewers or developers. We then identified five causes:

\begin{itemize}
    \item \textbf{Violation of coding conventions}: certain violations of coding conventions (e.g. naming convention) cause the smell. (Example: ``\emph{moreThanOneIp (CamelCase) is not our naming convention}'' \footnote{\url{https://review.opendev.org/\#/c/147739/}}).
    \item \textbf{Lack of familiarity with existing code}: developers introduced the smell due to unfamiliarity with the functionality or structure of the existing code. (Example: ``\emph{this useless line because \texttt{None} will be returned by default}'' \footnote{\url{https://review.opendev.org/\#/c/147042/}}).
    \item \textbf{Unintentional mistakes of developers}: the developer forgets to fix the smell or introduces the smell by mistake. (Example: ``\emph{You can see I renamed all of the other test methods and forgot about this one}'' \footnote{\url{https://review.opendev.org/\#/c/125384/}}).
    \item \textbf{Improper design}: the smell is identified to be related to improper design of the code. (Example: ``\emph{...If that's the case something is smelly (too coupled)...}'' \footnote{\url{https://review.opendev.org/\#/c/181674/}}).
    \item \textbf{Detection by code analysis tools}: the reviewer points out that the smell was detected by code analysis tools. (Example: ``\emph{\texttt{pass} is considered as dead code by python coverage tool}'' \footnote{\url{https://review.opendev.org/\#/c/143709/}}).
\end{itemize}

\begin{figure}[htb]
    \centering
    \includegraphics[width=\linewidth]{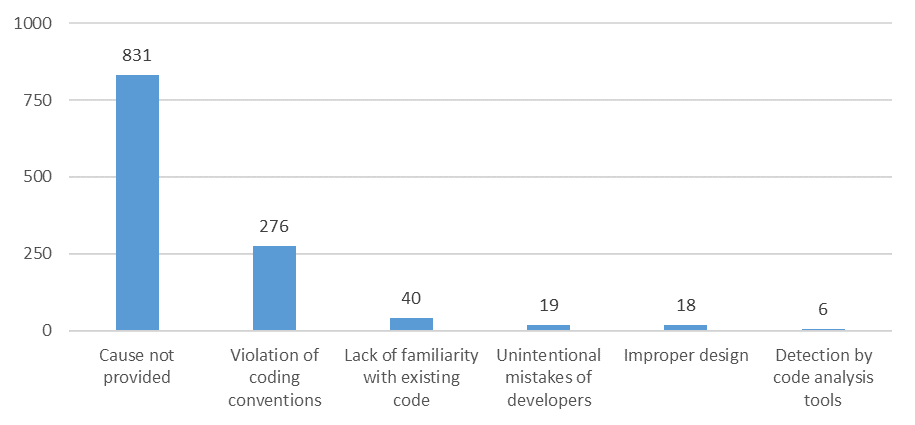}
    \caption{Reasons for the identified smells}
    \label{fig:reasons_for_smells}
\end{figure}

As demonstrated in Fig. \ref{fig:reasons_for_smells}, we found that the majority of reviews (70\%) did not provide any explanation for the identified smells - in most cases, the reviewer(s) simply pointed out the problem, but did not provide any further reasoning for their decisions. 
276 (23\%) of the reviews indicate that \textit{violation of coding conventions} is the main reason for the smell. For example, a reviewer suggested that the developer should adhere to the naming standard of `test\_[method under test]\_[detail of what is being tested]', as shown below:

\begin{tcolorbox}[colback = white]
\textbf{Link:} \url{http://alturl.com/urn8k} \\
\textbf{Reviewer:} ``Please adhere to the naming standard of `test\_[method under test]\_[detail of what is being tested]' to ensure that future maintainers will have an easier time associating tests and the methods they target.''
\end{tcolorbox}

In addition, 40 (3\%) of the reviews indicate that the smells were caused by developers' \textit{lack of familiarity with existing code}. An example of such a case is shown below. In this case, the reviewer pointed out that the exception handling should be removed. It could imply that the developer was not aware that the specific exception is not raised.

\begin{tcolorbox}[colback = white]
\textbf{Link:} \url{http://alturl.com/ccjy3} \\
\textbf{Reviewer:} ``on block\_device.BlockDeviceDict.from\_api(), exception.InvalidBDMVolumeNotBootable does not raise. so it is necessary to remove the exception here.''
\end{tcolorbox}

Nineteen reviews attributed \textit{unintentional mistakes of developers} (such as copy and paste) to be the cause of the smell, similar to the example shown below:

\begin{tcolorbox}[colback = white]
\textbf{Link:} \url{http://alturl.com/zwz2x}\\
\textbf{Reviewer:} ``I think you forgot to remove this.''\\
\textbf{Developer:} ``Darn, yes bad copy / paste.  Will fix it.''
\end{tcolorbox}

Eighteen reviews indicate that \emph{improper design} was the cause for the identified smell. 
In the rest (6) of the reviews, reviewers would note that the smell was detected by code analysis tools. For example, a reviewer pointed out that the code `pass' would be regarded as dead code by coverage tool.

\begin{tcolorbox}[colback = white]
\textbf{Link:} \url{http://alturl.com/azt42}\\
\textbf{Reviewer:} ``you can remove `pass', it's commonly considered as dead code by coverage tool''
\end{tcolorbox}

\noindent\resizebox{\columnwidth}{!}{\fbox{
	\parbox{\columnwidth}{
		\textbf{RQ2: common causes for smells as identified during code reviews.}
		\\
		Taken overall, over half of the reviews did not provide an explanation of the cause of the smells. In terms of the formulated causes, \textbf{\emph{violation of coding conventions}} is the main cause for the smells as noted by reviewers and developers.
	}
}}
\\

\textbf{RQ3: How do reviewers and developers treat the identified code smells?}

\textbf{RQ3.1: What actions do reviewers suggest to deal with the identified smells?}

The results of this research question are shown in Table \ref{tab:actions_reviewers}.
In the majority of reviews (870, representing 73\% of the reviews), reviewers recommended \emph{fix} for resolving the identified code smells. These fixes include either general directions (such as the name of a refactoring technique to be used) or specific actions (points to specific changes to the code base that could remove the smell). 303 (35\%) of these fixes provided example code snippets to help developers better refactor the smells.
Below is an example of a review that suggested a \emph{fix} recommendation. In this example, the reviewer suggested removing duplicated code from a test case, and also provided a working example of how to apply ``extract method'' refactoring to define a new test method, so that it could be referenced from multiple methods to remove code duplication.

\begin{tcolorbox}[colback = white]
\textbf{Link:} \url{http://alturl.com/c3g69}\\
\textbf{Reviewer:} ``I think you can do function that remove duplicated code, something like that following...''
\begin{lstlisting}[language=Python,breaklines=true,basicstyle=\small]
def _compare(self, exp_real):
    for exp, real in exp_real:
        self.assertEqual(exp['count'], real.count)
        self.assertEqual(exp['alias_name'], real.alias_name)
        self.assertEqual(exp['spec'], real.spec)
\end{lstlisting}
\end{tcolorbox}

272 reviews (23\%) fell under the \textit{capture} category. In those reviews, the reviewers just pointed to the presence of the smells, but did not provide any refactoring suggestions. In a small number of reviews (48, 4\%), reviewers suggested ignoring the code smell found in the code review.

\begin{table}[h]
\caption{Actions recommended by reviewers to resolve smells in the code}
\label{tab:actions_reviewers}
\resizebox{\columnwidth}{!}{%
\begin{tabular}{@{}l|l@{}}
\toprule
\textbf{Reviewer's recommendation}                          & \textbf{Count} \\ \midrule
Fix (without recommending any specific implementation)      & 567            \\
Fix (provided specific implementation)                      & 303            \\
Capture (just noted the smell)                              & 272            \\
Ignore (no side effects)                                    & 48             \\ \bottomrule
\end{tabular}}
\end{table}

\textbf{RQ3.2: What actions do developers take to resolve the identified smells?}

Table \ref{tab:fixed_smells} provides details of the number of reviews that identified code smells versus the number of fixes of the identified code smells.
Of the 1,190 code smells identified in the reviews, the majority (1,029, representing 86\%) were refactored by the developers after the review (i.e., changes were made to the patch). The remainder did not result in any changes in the code, indicating that the developers chose to ignore such recommendations. This could be a case where developers thought that those smells were not as harmful as suggested by the reviewers, or that there were other issues requiring more urgent attention, resulting in those smells being counted as technical debt in the code.

As per the results of RQ1, \emph{duplicated code}, \emph{bad naming}, and \emph{dead code} were the most frequently identified smells by reviewers. Those smells were also widely resolved by developers. Over 508 (82\%) \emph{duplicated code}, 276 (91\%) \emph{bad naming}, and 210 (95\%) \emph{dead code} smell instances were refactored by developers after they were identified in the reviews.
The proportion of other smells being fixed was nearly 78\% (35/45). However, the sample size for these smells (35 instances) is still too small to make any generalisations.

\begin{figure*}[ht]
    \centering
        \frame{\includegraphics[width=1\linewidth]{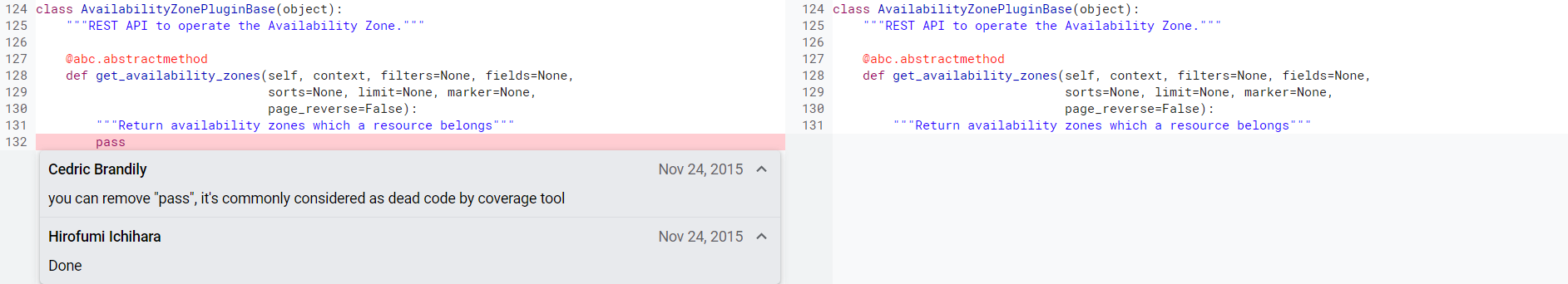}}

    \subfloat[\label{fig:deadcode:before} method before review.]{\hspace{.5\linewidth}}
    \subfloat[\label{fig:deadcode:after} after change made by the developer.]{\hspace{.5\linewidth}}
    \caption{An example of a \emph{remove dead code} operation after review \small (the change is highlighted in Line 132 (a))}
    \label{fig:deadcode_example}
\end{figure*}

Below is an example of a review with a recommendation by the reviewer to remove  \emph{dead code} in Line 132 of the original file (i.e., remove the \texttt{pass} statement); the developer then agreed to the reviewer's recommendation and deleted the unused code. Fig. \ref{fig:deadcode_example} shows the code before review (\ref{fig:deadcode:before}) and after the action taken by the developer (\ref{fig:deadcode:after}). 

\begin{tcolorbox}[colback = white]
\textbf{Link:} \url{http://alturl.com/szswu}\\
\textbf{Reviewer:} ``you can remove `pass', it's commonly considered as dead code by coverage tool''\\
\textbf{Developer:} ``Done''
\end{tcolorbox}

\begin{table}[h]
\caption{Developers' actions to code smells identified during reviews}
\label{tab:fixed_smells}
\resizebox{\columnwidth}{!}{%
\begin{tabular}{l|c|c|c}
\toprule
\textbf{Code smell  }   & \textbf{\#Reviews }       & \textbf{\#Fixed by developers}    & \textbf{\% of fixes }   \\ \midrule
Duplicated Code         & 620                       & 508                               & 82\%      \\
Bad Naming              & 304                       & 276                               & 91\%      \\
Dead Code               & 221                       & 210                               & 95\%      \\
Long Method             & 30                        & 25                                & 83\%      \\
Circular Dependency     & 3                         & 2                                 & 67\%      \\ 
Swiss Army Knife        & 1                         & 1                                 & 100\%     \\
General Smell           & 11                        & 7                                 & 64\%      \\ \hline
\textbf{Total}          & \textbf{1190}             &\textbf{1029}                      & \textbf{86\%}     \\ \bottomrule
\end{tabular}}
\end{table}

\textbf{RQ3.3: What is the relationship between the actions suggested by reviewers and those taken by developers?}

\begin{figure}[htb]
    \centering
    \captionsetup{justification=centering}
    \includegraphics[width=1\linewidth]{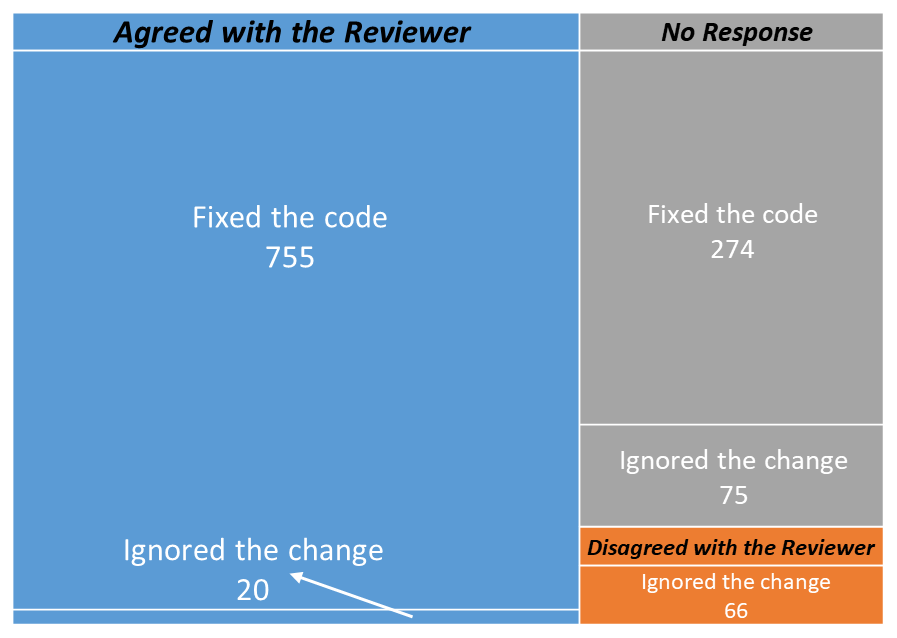}
    \caption{A treemap of the relationship between developers' actions in response to reviewers' recommendations regarding code smells identified in the code}
    \label{fig:actions_map}
\end{figure}

For answering this RQ, a map of reviewer recommendations and resulting developer actions is shown in Fig. \ref{fig:actions_map}. 
In 775 (65\%) of the obtained reviews, developers agreed with the reviewers' suggestions and took exactly the same actions (either \textit{fix} or \textit{ignore}) as suggested by reviewers. Of those cases, there are 20 cases where developers agreed with reviewers on ignoring the smell (i.e., a smell has been identified, but the reviewer may think that the impact of the smell is minor). The example below shows a case where a reviewer pointed out that they could accept duplicated code if there was a reasonable justification and the developer gave their explanation and ignored the smell.

\begin{tcolorbox}[colback = white]
\textbf{Link:} \url{http://alturl.com/s59so}\\
\textbf{Reviewer:} ``...I just don't like duplicated code but if there is a reasonable justification for this I can be sold cheaply and easily.'' \\
\textbf{Developer:} ``we need \texttt{create\_vm} here to support a lot of the other testing in this method. I agree it's duplicate code, but it's needed here too and this one is more complex that (sic) the \texttt{test\_config} one....''
\end{tcolorbox}

In 274 (23\%) reviews, even when developers did not respond to reviewers directly in the review system, they still made the required changes to the source code files. 
We noted another 66 (5\%) reviews where developers had different opinions from reviewers and decided to ignore the recommendations to refactor the code and remove the smell. In those cases, the developers themselves decided that the smells were either not as critical as perceived by the reviewers, or there are time or project constraints preventing them from implementing the changes, which is typically self-admitted technical debt \cite{potdar2014exploratory}. An example review is shown below:

\begin{tcolorbox}[colback = white]
\textbf{Link:} \url{http://alturl.com/pzmzz}\\
\textbf{Reviewer:} ``This method has a lot duplicated code of `\_apply\_instance\_name\_template'. The differ in the use of `index' and the CONF parameters. With a bit refactoring only one method would be necessary I guess.''\\
\textbf{Developer:} ``I thought to make / leave this separate in case one wants to configure the multi\_instance\_name\_template different to that of single instance.''
\end{tcolorbox}

Similarly, there were also 75 (6\%) reviews in which developers neither replied to reviewers nor modified the source code. For those cases, we assume that developers did not find the recommendations regarding how to deal with the specific smells in the code helpful, and therefore decided not to perform any changes. In all of those cases, no further explanation/reasons were provided by the developers on why they ignored these recommended changes.\\ 

\noindent\resizebox{\columnwidth}{!}{\fbox{%
	\parbox{\columnwidth}{%
		\textbf{RQ3: reviewers' recommendations and developers' actions.}
		\\
		In most reviews, reviewers provided fixing (refactoring) recommendations (e.g., in the form of code snippets) to help developers remove the identified smells. Developers generally followed those recommendations and performed the suggested refactoring operations, which then appeared in the patches committed after the review.
	}
}}

\section{Discussion}
\label{sec:discussion}

\subsection{RQ1: The most frequently identified smells}

In general, code smells are not commonly identified during code reviews. The results of RQ1 imply that \textit{duplicated code}, \textit{bad naming}, and \textit{dead code} were, by far, the most frequently identified code smells in code reviews. The results regarding \textit{duplicated code} are in line with previous findings which indicate that the smell is also frequently discussed among developers in online forums \cite{tahir2020stackexchange}, and is also the smell that developers are most concerned about \cite{Yamashita2013e}. However, \textit{dead code} and \textit{bad naming} were not found to be ranked highly in previous studies \cite{Yamashita2013e}. 

The different results are due to the different context and domain, which are critical in identifying smells, as shown by previous studies \cite{Yamashita2013e,tahir2020stackexchange}. The results reported in these two previous studies \cite{tahir2020stackexchange, Yamashita2013e} are based on more generic investigation of code smells among online Q\&A forums' users and developers. The context of some of these code smells was not fully taken into account, even if the developers may provide some specific scenario to explain their views. In contrast, our study is project-centric, and the context of the identified code smells during code reviews is known to reviewers and developers involved in identification and removal of the smells.

\subsection{RQ2: The causes for identified smells}
We identified five types of common causes for code smells in code reviews (RQ2). Among these, \emph{violation of coding conventions} is the major cause of code smells identified in reviews. Coding conventions are important in reducing the cost of software maintenance while the existence of smells can increase this cost. We conjecture that this is because that developers may not be familiar with the coding conventions of their community and the system they implemented. For example, \emph{duplicated code} and \emph{dead code} may occur because developers are not aware of existing functionality, while \emph{bad naming} may occur because developers are not familiar with the naming conventions. This reason can imply that developers can inadvertently violate coding conventions in their company or community, leading to smells or other problems. This may have a negative impact on software quality.

Another main observations is that more than half of reviewers (in review comments where they indicated that there was a code smell) simply pointed out the smell in the code, but did not provide any further explanation of why they considered that as a smell. One explanation for this is that the identified smells were simple or self-explanatory (e.g., \textit{duplicated code}, \textit{dead code}). Therefore, it is not expected that reviewers need to provide further explanation for these smells. Although the point of code review is to identify shortcomings (e.g., code smells) in the contributed code, understanding the causes of code smells can help practitioners understand how the code smell is introduced, and then take corresponding measures. 

\subsection{RQ3: The relationship between what reviewers suggest and the actions taken by developers}
The results of RQ3 show that reviewers usually provide useful recommendations (sometimes in the form of code snippets)  when they identify smells in the code and developers usually follow these suggestions. Given the constructive nature of most reviews, developers tend to agree with the review-based smell detection mechanism (i.e., where a reviewer detects and reports a smell) and in most cases they perform the recommended actions (i.e., refactoring their code) to remove the smell. We believe that this is because reviewers can take the contextual information into full account as the program context and domain are important in identifying smells \cite{Yamashita2013e,tahir2020stackexchange,sae2018context}.

Although not as frequent, there are cases where changes recommended by reviewers were ignored (see Figure~\ref{fig:actions_map}). This situation is partially due to the different understanding of reviewers and developers about the severity of identified code smells, i.e., when a reviewer identifies a code smell to be resolved, a developer may not agree that this code smell must be fixed, such as technical debt \cite{li2015systematic}.

\subsection{Implications}
First, although we built the initial set of keywords with 5 general code smell terms and 40 specific code smell terms, most of the smells were not identified in code reviews, such as \emph{long parameter list}, \emph{temporary field}, and \emph{lazy class}. One potential reason is that code smells which are considered as problematic in academic research may not be considered as a pressing problem in industry. More research should be conducted with practitioners to explore existing code smells and to understand the driving force behind industry efforts on code smell detection and elimination. This will further help guide the design of next-generation code smell detection tools.

Second, \emph{violation of coding conventions} is the main cause of code smells identified in code reviews. It implies that developers' lack of familiarity with the coding conventions in their company or organization can have a significantly negative impact on software quality. To reduce code smells, project leaders not only need to adopt code analysis tools, but also need to help and educate their developers to become familiar with the coding conventions adopted in the system.

Third, in smell-related reviews, reviewers usually give useful suggestions to help developers better fix the identified code smells and developers generally tend to accept those suggestions. It implies that review-based detection of smells is seen as a trustworthy mechanism by developers. In general, code reviews are useful for finding defects and locating code smells. Although code analysis tools (both static analyzers and dynamic (coverage-based) tools) are able to find some of those smells, their large outputs restrict their usefulness. 
Most tools are context and domain-insensitive, making their results less useful due to the potential false positives \cite{Fontana2016}. Context seems to matter in deciding whether a smell is bad or not \cite{tahir2020stackexchange,sharma2018survey}. There have been some recent attempts to develop smell-detection tools that take developers-context into account \cite{sae2018context,pecorelli2020developers}. Still, other contextual factors such as project structure and developer experience are much harder to capture with tools. Code reviewers are much better positioned to understand and account for those contextual factors (as they are involved in the project) and therefore their assessment of smells might be trusted more by developers than automated detection tools. 
To increase reliability, it may be that we need a two-step detection mechanism; static analysis tools to identify smells (as they are faster than human assessment and also scalable) and then for reviewers to go through those smell instances. They should decide, based on the additional contextual factors, which of those smells should be removed and at what cost. The problem with such an approach is that most tools would probably produce large sets of outputs, making it impractical for reviewers working on a large code base.
\section{Threats to Validity}
\label{sec:threats}

\textbf{External Validity}: our study considered two major projects from the OpenStack community (Nova and Neutron), since those projects have invested a significant effort in their code review process (see Section \ref{sec:subject_projects}). Due to our sample size, our study may not be generalizable to other systems. However, we believe that our findings could help researchers and developers understand the importance of the manual detection of code smells better. Including code review discussions from other communities will supplement our findings, and this may lead to more general conclusions.

\textbf{Internal Validity}: the main threat to internal validity is related to the quality of the selected projects. It is possible that the projects we included do not provide a good representation of the types of code smells we included in our study. While we only selected two projects from the OpenStack community with Gerrit as their code review tool, OpenStack investment in code review processes and commitment to perform code review to their entire code base and following coding best practices make it a good candidate for our analysis.

\textbf{Construct Validity}: a large part of the study depends on manual analysis of the data, which could affect the construct validity due to personal oversight and bias. In order to reduce its impact, each step in the manual analysis (i.e., identifying smell-related reviews and classification) was conducted by at least two authors, and results were always cross-validated. The selection of the keywords used to identify the reviews which contain smell discussions is another threat to construct validity since reviewers and developers may use terms other than those that we used in our mining query. To minimize the impact of this threat, we first combined a list of code smell terms that developers and researchers frequently used, as reported in several previous studies. Then, we identified the keywords by following the systematic approach used by Bosu \textit{et al.} \cite{bosu2014identifying} to minimize the impact of missing keywords due to misspelling or other textual issues. Moreover, we randomly selected a collection of review comments that did not contain any of our keywords to supplement our approach, reducing the threat to the construct validity.

\textbf{Reliability}: before starting our full scale study, we conducted a pilot run to check the suitability of the data source. Besides, the execution of all the steps in our study, including the mining process, data filtering, and manual analysis, was discussed and confirmed by at least two of the authors.

\section{Conclusions}
\label{sec:conclusion}
Code review is a common software quality assurance practice. One of the issues that may impact software quality is the presence of code smells. Yet, little is known about the extent to which code smells are identified and resolved during code reviews. To this end, we performed an empirical study of code smell discussions in code reviews by collecting and analyzing code review comments from the two most active OpenStack projects (Nova and Neutron). Our results show that: 1) code smells are not commonly identified in code reviews, and when identified, \emph{duplicated code}, \emph{bad naming}, and \emph{dead code} are, by far, the most frequently identified smells; 2) \emph{violation of coding conventions} is the most common cause for smells as identified during code reviews; 3) when smells are identified, most reviewers provide recommendations to help developers fix the code and remove the smells (via specific refactoring operations or through an example code snippet); and 4) developers mostly agree with reviewers and remove the identified smells through the suggested refactoring operations.

Our results suggest that: 1) developers should follow the coding conventions in their projects to reduce code smell incidents; and 2) code smell detection via code reviews is seen as a trustworthy approach by developers (given their constructive nature) and smell-removal recommendations made by reviewers appear more actionable by developers. We found that the majority of smell-related recommendations were accepted by developers. We believe this is mainly due to the context-sensitivity of the reviewer-centric smell detection approach.

We plan to extend this work by studying code reviews in a larger set of projects from different communities. We also plan to explore, in more detail, the refactoring actions taken by developers when removing certain smells, and the reasons (e.g., trade-off in managing technical debt \cite{li2015systematic}) why developers disagreed with reviewers' recommendations or ignored the recommended changes by e.g., code smell discussions \cite{shcherban2020aic}. 

\balance
\bibliographystyle{ieeetr}
\bibliography{references}

\end{document}